\documentclass[aps,prl,twocolumn,amsmath,amssymb,superscriptaddress,showpacs,10pt]{revtex4-1}

\usepackage{graphicx}
\usepackage{mathrsfs}
\usepackage{bm}
\usepackage[colorlinks]{hyperref}
\begin{document}

\newcommand{\ms}[1]{\mbox{\scriptsize #1}}
\newcommand{\msb}[1]{\mbox{\scriptsize $\mathbf{#1}$}}
\newcommand{\msi}[1]{\mbox{\scriptsize\textit{#1}}}
\newcommand{\nn}{\nonumber} 
\newcommand{\dg}{^\dagger}
\newcommand{\smallfrac}[2]{\mbox{$\frac{#1}{#2}$}}
\newcommand{\pfpx}[2]{\frac{\partial #1}{\partial #2}}
\newcommand{\dfdx}[2]{\frac{d #1}{d #2}}
\newcommand{\half}{\smallfrac{1}{2}}
\newcommand{\s}{{\mathcal S}}
\newcommand{\red}{\color{red}}
\newcommand{\bluetext}{\color{blue}}
\newcommand{\rws}{\color{blue}}
\newcommand{\kurt}{\color{green}} 
\newtheorem{theo}{Theorem} \newtheorem{lemma}{Lemma}
\def\r{\hat{G}^R}
\def\a{\hat{G}^A}
\def\k{\hat{G}^K}
\def\gs{\check{g}_s}
\def\gp{\check{g}_p}
\def\qr{\hat{g}^R}
\def\qa{\hat{g}^A}
\def\qk{\hat{g}^K}
\def\der{\partial_{\vec{R}}}
\def\dif{\partial_{'}}

\renewcommand{\theequation}{\thesection\arabic{equation}}

\title{Nonlocal superconducting quantum interference device }
\author{Taewan Noh}
\altaffiliation{Present address:  National Institute of Standards and Technology (NIST), Boulder, CO, USA}
\address{Department of Physics, Northwestern University, Evanston, Illinois. 60208, USA}
\author{Andrew Kindseth}
\address{Department of Physics, Northwestern University, Evanston, Illinois. 60208, USA}
\author{Venkat Chandrasekhar}
\email{Email address: v-chandrasekhar@northwestern.edu}
\address{Department of Physics, Northwestern University, Evanston, Illinois. 60208, USA}

\begin{abstract}
Superconducting quantum interference devices (SQUIDs) that incorporate two superconductor/insulator/superconductor (SIS) Josephson junctions in a closed loop form the core of some of the most sensitive detectors of magnetic and electric fields currently available.  SQUIDs in these applications are typically operated with a finite voltage which generates microwave radiation through the ac Josephson effect.  This radiation may impact the system being measured.  We describe here a SQUID in which the Josephson junctions are formed from strips of normal metal (N) in good electrical contact with the superconductor (S).  Such SNS SQUIDs can be operated under a finite voltage bias with performance comparable or potentially better than conventional SIS SQUIDs.  However, they also permit a mode of operation that is based on the unusual interplay of quasiparticle currents and supercurrents in the normal metal of the Josephson junction.  The method allows measurements of the flux dependence of the critical current of the SNS SQUID without applying a finite voltage bias across the SNS junction, enabling sensitive flux detection without generating microwave radiation.    
\end{abstract} 
\date{\today}


\maketitle

\section{I. Introduction}

Superconducting quantum interference devices (SQUIDs) are the most sensitive flux detectors available, and have found widespread use in various fields. The most widely used type of SQUID is the dc SQUID, consisting of two Josephson junctions connected in parallel to form a loop \cite{clarke,braginski,vanduzer}, with flux sensitivities better than $10^{-6} \Phi_0/\sqrt{\text{Hz}}$ now fairly common ($\Phi_0 = h/2e = 2.07 \times 10^{-15}$ T m$^2$ is the superconducting flux quantum) \cite{braginski,carelli,fagaly}.    The dc SQUID is typically operated in the finite voltage regime,  biased with a current larger than the critical current $I_c$ of the SQUID.  In this mode, the voltage across the SQUID is a periodic function of the magnetic flux $\Phi$ through the  SQUID loop.  However, using a dc SQUID with a finite voltage bias not only generates a small amount of dissipation, it also generates radiation through the ac Josephson effect.  This radiation may affect the sample, for example by causing heating, or through the direct effect of microwaves on quasiparticle excitations in the sample \cite{jayich}. Thus, a SQUID that can operate without a finite voltage bias across the Josephson junctions is of interest.  SQUIDs can be operated in the so-called dispersive mode \cite{hatridge} without a finite voltage.  However, such devices still use microwaves for operation, leaving the possibility that the microwave drive will affect the sample being measured.    

SQUIDs typically incorporate superconductor/insulator/superconductor (SIS) junctions, where the two superconductors are separated by a thin ($\sim$2 nm) insulating tunnel barrier.  A variety of different `weak links' can replace the insulator in a SIS junction to form different types of Josephson junctions \cite{likharev}, but with the exception of superconductor/normal-metal/superconductor (SNS) junctions, it is difficult to fabricate practical dc SQUIDs with these other technologies.  SNS dc SQUIDs can also be operated in a finite voltage bias mode analogous to conventional SIS dc SQUIDs \cite{angers}. However, SNS SQUIDs can also be operated in modes where all superconducting elements are at the same potential.  For example, one can detect the flux coupled into the SQUID loop by detecting the modulation of the quasiparticle density of states in the normal part of a SNS junction with the coupled flux \cite{giazotto,jabdaraghi}.  We show here that the unusual interplay of quasiparticles and supercurrents in the normal part of SNS junctions enables an entirely different mode of operation, where $I_c(\Phi)$ can be detected by a simple resistance measurement even when the voltage between the two superconductors of the SQUID remains zero, and thus no Josephson radiation is generated.

Supercurrent flow between the two superconductors in a SNS junction is enabled by the superconducting proximity effect induced in the normal metal \cite{deutscher}.  In the diffusive limit \cite{usadel}, the upper limit to the length $L$ of the normal metal in the SNS junction is set by two length scales:  the electron phase coherence length $L_\phi$ and the thermal diffusion length $L_T = \sqrt{\hbar D/k_B T}$,  where $D$ is diffusion coefficient of electrons in the normal metal and $T$ is the temperature \cite{chandra2}.  To obtain a significant supercurrent, $L$ should be much shorter than both $L_\phi$ and $L_T$.  Since $L_\Phi$ and $L_T$ can be many microns at millikelvin, one can fabricate extended SNS junctions with micron size dimensions.

\begin{figure*}
\includegraphics[width=18cm]{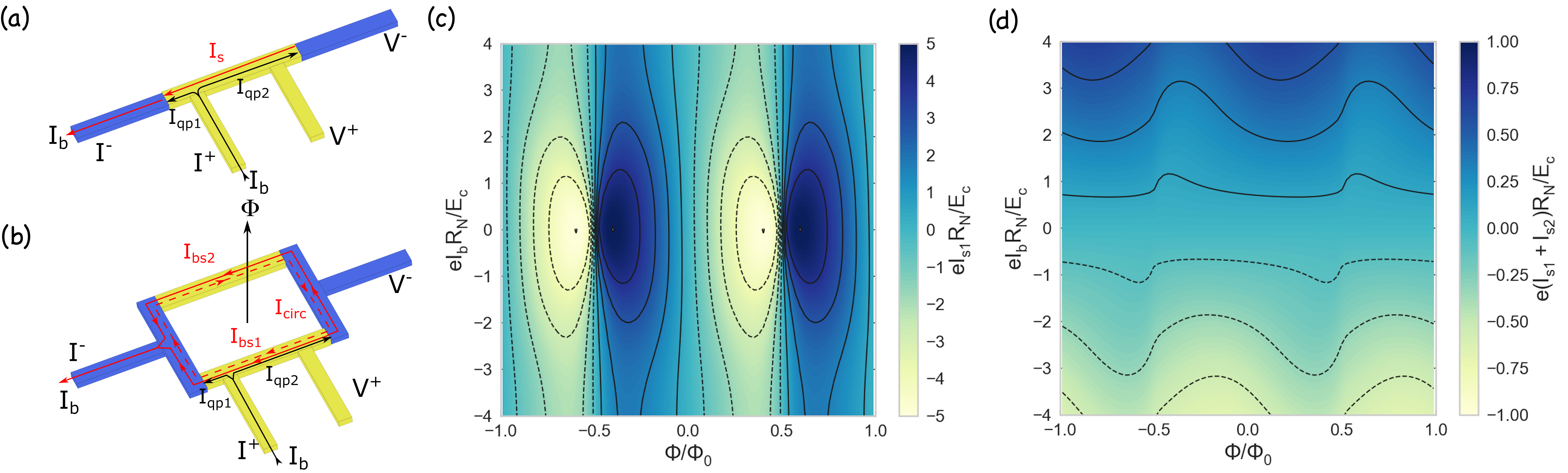}
\caption{(a)  Schematic of a symmetric SNS Josephson junction with multiple normal metal leads.  Yellow (blue) represents normal metal (superconductor).  (b)  Nonlocal SNS SQUID.  (c)  Total supercurrent $I_{s1}=I_{bs1} + I_{circ}$ in the multiterminal junction as a function of $\Phi$ and injected current $I_b$. The critical current at $I_b=0$ is given by the scale of the plot $\sim 5 E_c/eR_N$. (d)  Sum of the supercurrents in both junctions.  The superconducting gap $ \Delta = 63.6 E_c$ and temperature $T = 0.8 E_c/k_B$ for these simulations. }
\label{fig1}
\end{figure*}

\section{II. Principle of operation}

The extended nature of the SNS junction allows one to place additional normal metal contacts on the normal part of the junction, enabling a mode of operation not possible with conventional SIS dc SQUIDS.  To see this, consider first an isolated SNS junction with two additional normal metal probes, as shown in Fig. \ref{fig1}(a).  A small transport current $I_b$ is sourced through one normal lead ($I^+$) and drained through a superconducting contact ($I^-$).  The two superconductors are Josephson coupled and at the same potential.  Consequently, the injected quasiparticle current splits into two branches ($I_{qp1}, I_{qp2}$), one branch going to each superconducting contact.  However, the second superconductor is a voltage probe ($V^-$), so that no net current can flow into it.  The quasiparticle current $I_{qp2}$ is therefore converted into a supercurrent $I_s$ at the NS interface that counterflows back to the first superconductor \cite{crosser,clarke2,tinkham,schmid}.  A \textit{nonlocal} voltage $V_{nl}$ develops between the second normal contact ($V^+$) and $V^-$ due to $I_{qp2}$ which is approximately $I_{qp2} R$, where $R$ is the resistance of the normal metal between $V^+$ and $V^-$. 
A dissipationless supercurrent between the two superconductors implies a phase difference $\Delta \phi$ between them.  This phase difference modifies the resistance of the normal metal through the proximity effect \cite{petrashov,nazarov}, giving rise to a variation of the differential resistance $dV_{nl}/dI_b$ with increasing $I_b$, with the maximum variation in resistance of order 10\% with perfectly transparent NS interfaces.  $I_s$ increases with $I_b$; at some point, $I_s$ exceeds $I_c$, the two superconductors are no longer at the same potential, and the nonlocal resistance abruptly drops.  This behavior has been verified experimentally \cite{noh}.  

Now consider two such SNS junctions in a dc SQUID, with additional normal metal leads attached to one, as shown in Fig. \ref{fig1}(b).  As before, for low bias currents, the source current $I_b$ will split into two quasiparticle currents $I_{qp1}$ and $I_{qp2}$.  $I_{qp2}$ will again be converted to supercurrent at the NS interface.  However, there are now two possible paths for this supercurrent to return to the current drain $I^-$.  One path is through the same junction ($I_{s1}$) and the other path is through the second junction ($I_{s2}$) with the requirement that $I_{qp2}= I_{s1} + I_{s2}$.  As before, $I_{s1} + I_{s2}$ will increase with increasing $I_b$, resulting in an increasing phase difference between the two superconductors, and a consequent modulation of the nonlocal differential resistance as in the linear structure.  However, we can now also thread a flux $\Phi$ through the SQUID loop; $\Phi$ will result in an additional circulating supercurrent $I_{\rm{circ}}$.  The situation is similar to a dc SQUID measured in the conventional manner with $I_{qp2}$ taking the place of the bias current $I_b$, with the important difference that we can measure a finite nonlocal differential resistance even when the voltage difference between the superconductors is still zero.  As we shall see below, this capability allows us to determine $I_c$ of the device without a voltage drop across the superconductors.

\begin{figure*}
\includegraphics[width=18cm]{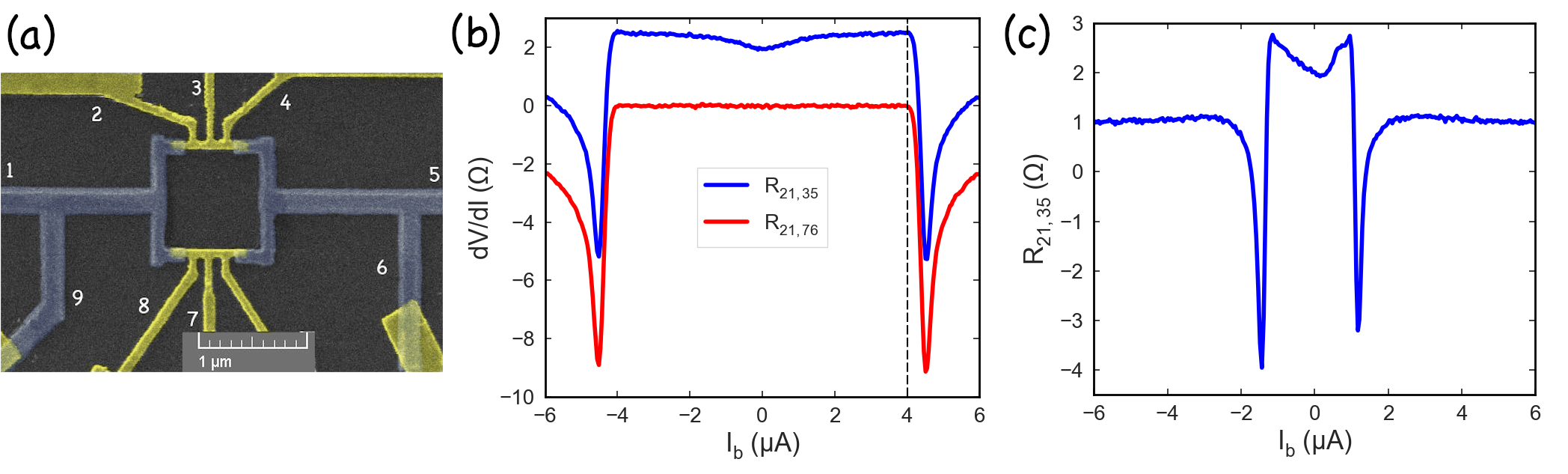}
\vspace{-0.3cm}
\caption{(a)  False color SEM image of a SNS SQUID.  Yellow represents the normal metal (Au) and blue represents the superconductor (Al).  Numbers identify contacts used in the four-terminal measurements.  (b) Measured nonlocal differential resistance $R_{21,35}$ (blue) and $R_{21,76}$ (red) as a function of the bias current $I_b$ applied between contacts 2 and 1.  A perpendicular magnetic field corresponding to a flux $-\Phi_0$ through the SQUID loop is applied.  (c)  $R_{21,35}$ measured with a flux $\Phi=0.45 \Phi_0$ through the SQUID loop.  All data taken at 26 mK. }
\label{fig2}
\end{figure*}

\section{III. Numerical Simulations}

To visualize how the flow of supercurrents and the nonlocal differential resistance varies as one changes $\Phi$, we have modeled the geometry of Fig. \ref{fig1}(b) using the quasiclassical equations of superconductivity in the diffusive limit \cite{usadel,chandra2}.  The simulations were done by solving the Usadel equations of quasiclassical superconductivity \cite{usadel} in the Riccati parametrization simultaneously with the kinetic equations for the quasiparticle distribution functions using open-source code \cite{virtanen,virtanen2} using the geometry of Fig. \ref{fig1}(b).  The details of the parametrization and code can be found in Appendix A, following Ref. [\citenum{virtanen3}].   For the simulations, a voltage $V_b$ was applied to the current bias lead, and the gauge invariant phase $\gamma_1$ and the voltage on the nonlocal lead $V_{nl}$ was varied iteratively in a numerical solver to satisfy two conditions:  (1) the current into the nonlocal voltage probe vanished, and 
(2) $I_{qp2}=I_{s1} + I_{s2}$.  The gauge invariant phase difference across the second SNS junction $\gamma_2$ is related to $\gamma_1$ by $\gamma_1 - \gamma_2 = 2 \pi \Phi/\Phi_0$.  The calculation was repeated for different values of $V_b$ and $\Phi$.  $I_b$ and $dV_{nl}/dI_b$ were then calculated numerically.  In order to keep the calculations tractable, the simulations assume perfect NS interface transparency and no voltage drop between the two superconductors.  Further details of the numerical simulations can be found in Appendix A \cite{dolgirev}.

Figure \ref{fig1}(c) shows the total supercurrent $I_{s1}=I_{bs1}+ I_{circ}$ in the multiterminal junction as a function of $\Phi$ and $I_b$.  $I_{s1}$ oscillates with $\Phi$ with a fundamental period of $\Phi_0$, with the amplitude of the oscillations being maximum for $I_b=0$ and decreasing with increasing $|I_b|$.  The supercurrent through the second junction $I_{s2}$ has similar behavior (not shown), except that the oscillations in $I_{s2}$ are 180$^\circ$ out of phase with the oscillations in $I_{s1}$.  The amplitude of the oscillations in $I_{s1}$ and $I_{s2}$ also differ slightly.  This difference arises from the difference in geometry between the two junctions, and the fact that a quasiparticle current is injected into the first junction, changing the quasiparticle distribution function and hence the supercurrent \cite{baselmans}.

Figure \ref{fig1}(d) shows the sum of the supercurrents in the two junctions $I_{s1} + I_{s2}$ as a function of $I_b$ and $\Phi$. The total supercurrent at a specific bias current $I_b$ oscillates as a function of $\Phi$, being in general larger when $\Phi \sim n \Phi_0$ and smaller when $\Phi \sim (n+1/2) \Phi_0$, where $n$ is an integer. In contrast to a SIS dc SQUID, the maxima and minima of the supercurrent do not occur exactly at $\Phi=n \Phi_0$ and $\Phi = (n+1/2) \Phi_0$ respectively.  For this SNS SQUID, there is an offset from these values that increases with increasing $|I_b|$ due to the aforementioned asymmetry in the junctions. While the simulations are performed assuming no voltage drop between the two superconductors so that we cannot determine $I_c$ directly, it can be seen that the supercurrent and hence $I_c$ oscillate as a function of $\Phi$.

\section{IV. Experimental Results}

\begin{figure*}
\includegraphics[width=18cm]{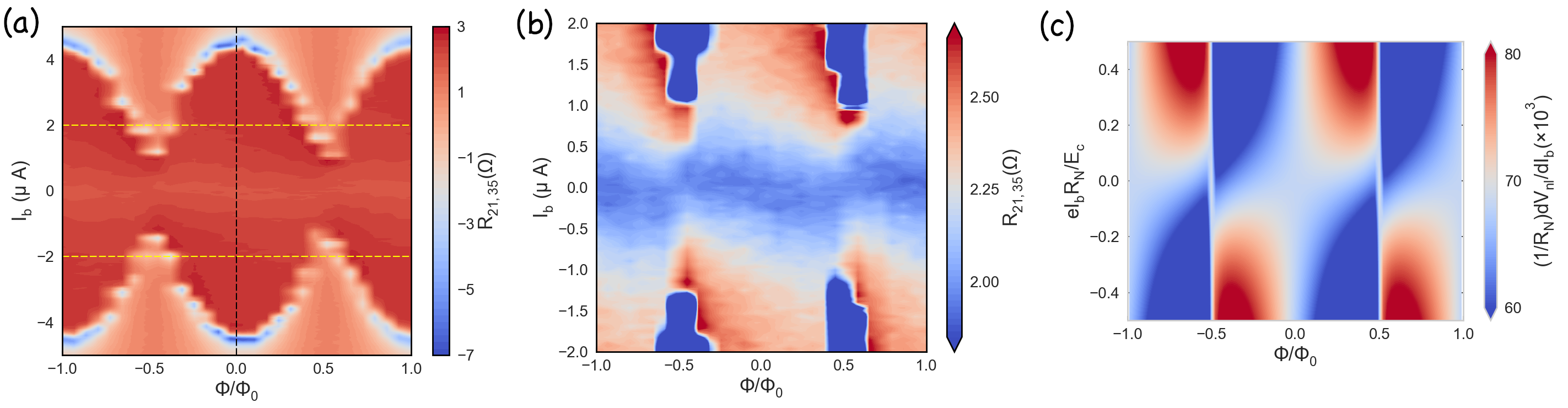}
\vspace{-0.6cm}
\caption{(a) Measured nonlocal differential resistance $R_{21;35}$ as a function of the bias current $I_b$ at various values of the flux $\Phi$ through the SQUID loop in the step of $\Phi/\Phi_0 =0.1$.  The dashed vertical line at $\Phi=0$ is provided to emphasize the slight asymmetry of the modulation with increasing $I_b$.  (b)  Expanded view of the data corresponding to the dashed yellow lines in (a).  Data taken at 26 mK.(c)  Numerical simulations of the nonlocal differential resistance $dV_{nl}/dI_B$ of the geometry of Fig. \ref{fig1}(b).   }
\label{fig3}
\end{figure*}

To demonstrate that these oscillations in $I_c$ can be detected without generating a finite voltage drop between the two superconductors, we fabricated and measured SNS loops with different geometries with Al as the superconductor and Au as the normal metal.  The devices were patterned using standard photolithography and multilevel electron-beam lithography techniques.  Prior to the Al deposition, the Au surface was cleaned with an \textit{in situ} argon ion etch to obtain clean interfaces between the Au and the Al.    The devices were loaded into an Oxford dilution refrigerator and cooled to 77 K within a few hours of the final deposition to preserve the quality of the Au/Al interface.
Four terminal resistance measurements in perpendicular magnetic fields were carried out using custom-built current sources that could provide both ac and dc currents simultaneously.  ac measurements were carried out using PAR 124 lock-in amplifiers at low frequencies (10s of Hz), with inputs to the lock-in amplifiers provided by custom-built, battery-operated low noise amplifiers housed in a mu-metal shield to avoid line frequency interference.  ac excitation currents of the order of 50 nA were used to avoid device heating due to the measurement.  

Figure \ref{fig2}(a) shows a false color SEM image of one of the devices with a geometry similar to that of Fig. \ref{fig1}(b), except that the normal sections of both SNS junctions have additional N leads attached and so are nominally identical.  An example of one of the other device geometries we fabricated and measured is shown in Appendix C.  The multiterminal nature allows us to measure various four-terminal differential resistances. Consequently, we use the common notation for the four-terminal resistance $R_{ij,kl}$, where the notation $R_{ij,kl} = dV_{kl}/dI_{ij}$ indicates that $i,j$ are the contacts where the ac current $I$ is sourced and drained, and $k,l$ are the contacts across which the resulting ac voltage drop is measured.

The blue curve in Fig. \ref{fig2}(b) shows the nonlocal differential resistance  $R_{21,35}$ as a function of $I_b$ at $T=26$ mK in a magnetic field corresponding to a flux $-\Phi_0$ through the area of the SQUID loop.  As schematically presented in Fig. \ref{fig1}(b), the nonlocal resistance arises from the quasiparticle current $I_{qp2}$ flowing through the N part of the junction. $R_{21,35}$ is approximately 2 $\Omega$ at $I_b=0$, rising symmetrically by about 10\% as $|I_b|$ is increased.  The resistance increase is due to the increasing phase difference between the two superconductors induced by the fraction $I_{s1}$ of the counterflowing supercurrent which modifies the resistance of the N part due to the proximity effect.  The remaining fraction $I_{s2}$ of the counterflowing supercurrent flows through the second SNS junction.  Since there is no quasiparticle current through the second SNS junction, there is no voltage drop between a N contact on this junction and either superconductor 
as shown in the nonlocal resistance  $R_{21,76}$ measured simultaneously. 

As $|I_b|$ is increased beyond $\sim 4$ $\mu$A, $R_{21,35}$ shows a sharp decrease, going to negative values of differential resistance when the counterflowing supercurrent in the device exceeds $I_c$, at which point the two superconductors are no longer at the same potential, so that the quasiparticle current $I_{qp2}$ drops.  Unlike the linear SNS junction $I_{qp2}$ does not vanish as there is still a path for the current to flow to the drain contact through the second SNS junction as a quasiparticle current \cite{noh}.  Consequently, $R_{21,76}$ also shows a sharp drop at the same values of $I_b$.  As $|I_b|$ is increased further, both nonlocal differential resistances approach their normal state values (modulated by the current distribution of the quasiparticle currents), positive for $R_{21,35}$ and negative for $R_{21,76}$ due to the relative orientation of the their respective voltage leads. Thus the maximum in $R_{21,35}$ gives a measure of the critical current $I_c$ of the SNS SQUID [indicated by the dashed line in Fig. \ref{fig2}(b)] while it is still in the zero voltage state.

To show that $I_c$ determined by this nonlocal measurement oscillates with applied flux as one expects in a dc SQUID, Fig. \ref{fig3}(a) shows $R_{21,35}$ as a function of $I_b$ and the normalized flux $\Phi/\Phi_0$ through the SQUID loop.  $I_c$ oscillates with $\Phi$ with a period of $\Phi_0$, varying from $\sim 4$ $\mu$A at $\Phi =0$ to $\sim1$ $\mu$A at $\Phi/\Phi_0=\pm 1/2$.  For an ideal SIS dc SQUID with identical Josephson junctions, one expects complete suppression of $I_c$ at $\Phi = (n+1/2)\Phi_0$.  In real SIS dc SQUIDs, differences between the critical currents of the two junctions will reduce the modulation in $I_c$ \cite{vanduzer}.  While the two SNS junctions in our device are nominally identical, the finite quasiparticle current in one junction results in a small difference in critical current between the two junctions \cite{baselmans}, resulting in a slight asymmetry in the interference pattern seen in Fig. \ref{fig3}(a) which increases with increasing $|I_b|$.  The asymmetry can be seen more clearly if we focus on the low bias regime $|I_b|<2$ $\mu$A, shown in Fig.  \ref{fig3}(b).  Numerical simulations of the nonlocal differential resistance of the schematic device of Fig. \ref{fig1}(b) shown in Fig. \ref{fig3}(c) exhibit the same qualitative asymmetric behavior, although the asymmetry is much more pronounced.  This is because the two SNS junctions in the simulated geometry of Fig. \ref{fig1}(b) are quite dissimilar.    

Operation of a dc SQUID in the conventional finite voltage bias mode involves biasing the SQUID with a modulation coil at a value of flux where the change in voltage $V$ with external flux ($dV/d\Phi$) is maximum, typically at $(n+1/4) \Phi_0$ \cite{vanduzer}. In an open loop configuration, the flux sensitivity of the SQUID has a lower limit determined by the intrinsic Johnson voltage noise of the SQUID $S_v = \sqrt{4 k_B T R}$ volts per unit bandwidth, i.e., $S_\Phi = S_v/(dV/d\Phi)$ \cite{fagaly}.  For the operation of our device as a flux sensor with no voltage drop between the two superconductors, we need to current bias the device.  The sensitivity of the device is then determined by the variation of the critical current $I_c$ with flux $dI_c/d\Phi$ and the intrinsic Johnson current noise $S_I = \sqrt{4 k_B T/R} \; \text{(Hz)}^{-1/2}$, $S_\Phi = S_I/(dI_c/d\Phi)$.  From Fig, \ref{fig3}(a), the maximum slope $dI_c/d\Phi$ occurs around $\Phi/\Phi_0 \sim 0.45$, where its value is $\sim 10$ $\mu$A/$\Phi_0$.  Assuming a resistance $R\sim 2$ $\Omega$ at $T=50$ mK, the expected flux noise of our device operated in the zero-voltage mode is $\sim 2 \times 10^{-7}\; \Phi_0/\sqrt{\text{Hz}}$.  Of course, with amplifiers and flux feedback schemes, the actual noise will be larger, but these numbers are comparable to conventional dc SIS SQUIDs \cite{fagaly}.

To use the device as a flux sensor, we need to be able to measure $I_c$ in the zero voltage state.  This can be done by using a feedback mechanism to change $I_b$ so that the nonlocal resistance is a maximum. Referring to Fig \ref{fig2}(b), this would be at $I_b\sim 4$ $\mu$A.  One way to do this is to use the measured $d^2 V/dI_b^2$ as the error signal for a current biasing feedback loop.  At the maximum in $dV/dI_b$, $d^2V/dI_b^2$ is zero and has opposite signs on either side of the maximum, and hence can in principle serve as an error signal.  Unfortunately, the nonlocal resistance trace in Fig. \ref{fig2}(b), which corresponds to an integral flux $n \Phi_0$ through the loop, has a rather broad maximum, making it difficult to maintain the device at $I_c$.  However, if we flux bias the device so that $dI_c/d\Phi$ is a maximum, as we would do in any case for maximum flux sensitivity, the maximum in the nonlocal resistance becomes much sharper, enhancing its suitability for feedback purposes.  This is demonstrated in Fig. \ref{fig2}(c), which shows the nonlocal differential resistance $R_{21,35}$ of the same device at $\Phi = 0.45 \Phi_0$.  While $I_c$ is now reduced, the peak in $R_{21,35}$ is much more pronounced, and $d^2V/dI_b^2$ about this point will show a much sharper slope and consequently serve as a much better error signal input to a feedback circuit. 

\section{V. Conclusion}

In summary, we have demonstrated the possibility of a mode of operation of SNS dc SQUIDs that uses the nonlocal resistance arising from the superconducting proximity effect to detect $I_c(\Phi)$ with no voltage drop between the two superconductors of the SQUID.  The Al/Au devices here were measured at millikelvin temperatures.  The limiting factor for higher temperature operation is the maximum possible length $L$ of the junction, whose value is determined by the condition $L>L_T$.  $L_T$ can in principle be made sufficiently long with very clean normal metals.  With Nb as the superconductor, such a device could then be operated at 4 K. \\

\section{Acknowledgments}

This work was partially supported by the US NSF under Grant No. DMR-1006445.  A.K. was partially supported by an Undergraduate Research Grant from the Weinberg College of Arts and Sciences, Northwestern University.

\appendix

\section*{Appendix A: Details of numerical simulations}
\subsection{1. Riccati parametrization of the quasiclassical equations of superconductivity}
\renewcommand{\thesection}{A}

The simulations in this paper numerically solve the quasiclassical equations of superconductivity in the diffusive limit in the Keldysh formulation, which gives both the Usadel equation for the quasiclassical superconducting Green's function as well as the equations for the quasiparticle distribution functions.  These are solved simultaneously for a network of one-dimensional (1D) normal metal wires that are connected to superconducting and normal metal ``reservoirs'', where the Green's function and the distribution function have well defined values.  
 
The equation for the retarded superconducting Green's function $\qr_s$, which is a solution of the Usadel equation, is given by
\begin{equation}
[\tau^3E +\tilde{\Delta},\tilde{g}_s^R] = iD\partial_{\vec{R}}(\tilde{g}_s^R \partial_{\vec{R}}\tilde{g}_s^R),
\label{eqn2.1}
\end{equation}
where $\tau^3$ is the usual Pauli spin matrix
\begin{equation}
\tau^3=
\begin{pmatrix}
1 & 0 \\
0 & -1 
\end{pmatrix},
\label{eqn2.2}
\end{equation}
and $\tilde{\Delta}$ is given by 
\begin{equation}
\tilde{\Delta} =
\begin{pmatrix}
0 & \Delta \\
-\Delta^* & 0
\end{pmatrix}.
\label{eqn2.3}
\end{equation}

The Green's function can be parametrized in different ways.  The open-source code used in this work uses the so-called Riccati parametrization, which is implemented in slightly different ways by different authors \cite{eschrig,hammer,cherkez}.  The code uses the formulation by Virtanen \cite{virtanen}, in which the retarded Green's function is expressed in terms of the complex Riccati parameters $\gamma$ and $\tilde{\gamma}$ as
 \begin{equation}
\qr_s = \frac{1}{1 + \gamma \tilde{\gamma}}   
\begin{pmatrix}
1-\gamma \tilde{\gamma} & 2 \gamma \\
2 \tilde{\gamma} & \gamma \tilde{\gamma} -1 )
\end{pmatrix}.
\label{eqn2.9}
\end{equation}
with the normalization condition $(\qr_s )^2=1$.  Putting this into the Usadel equation (\ref{eqn2.1}), we obtain from the off-diagonal components the following coupled equations for $\gamma$ and $\tilde{\gamma}$ in the normal metal wires (where $\Delta=0$):
\begin{subequations} \label{eqn2.10}
\begin{align}
D \left[\der^2 \gamma - \frac{2 \tilde{\gamma}}{1 + \gamma \tilde{\gamma}} (\der \gamma)^2 \right] +  2 i E \gamma &= 0, \label{eqn2.10a} \\
\intertext{and}
D \left[\der^2 \tilde{\gamma} - \frac{2 \gamma}{1 + \gamma \tilde{\gamma}} (\der \tilde{\gamma})^2 \right] +  2 i E \tilde{\gamma} &= 0. \label{eqn2.10b} 
\end{align}
\end{subequations}

\subsection{2. Spectral quantitites in terms of the Riccati parameters}

In terms of the Riccati parameters, the spectral supercurrent $Q$ is defined as
\begin{equation}
Q = 2 {\rm{Re}} \left[ \frac{1}{(1+ \gamma \tilde{\gamma})^2} (\gamma \der \tilde{\gamma} - \tilde{\gamma} \der \gamma) \right].
\label{eqn2.11}
\end{equation}
Here ${\rm{Re}}$ stands for the real part.

The modified diffusion coefficients $M_{ij}$ are given by
\begin{subequations} \label{eqn2.12}
\begin{align}
M_{00} & = \frac{1}{|1 + \gamma \tilde{\gamma}|^2} \left[(|\gamma |^2 - 1)(|\tilde{\gamma}|^2 -1) \right], \label{eqn2.12a}\\
M_{33} & = \frac{1}{|1 + \gamma \tilde{\gamma}|^2} \left[(|\gamma |^2 + 1)(|\tilde{\gamma}|^2 +1) \right], \label{eqn2.12b}\\
M_{03} & = \frac{1}{|1 + \gamma \tilde{\gamma}|^2} \left[|\tilde{\gamma}|^2 -|\gamma |^2 \right], \label{eqn2.12c}\\
\intertext{and}
M_{30} & = -\frac{1}{|1 + \gamma \tilde{\gamma}|^2} \left[|\tilde{\gamma}|^2 -|\gamma |^2 \right] = -M_{03}. \label{eqn21.2d}
\end{align}
\end{subequations}
The charge current $j(R,T)$ and thermal current $j_{th}(R,T)$ are given in terms of these quantities by
\begin{subequations} \label{eqn2.13}
\begin{align}
&j(R,T) \nonumber\\ 
&= eN_0 D \int dE \; [M_{33} (\partial_R h_T) + Q h_L + M_{03} (\partial_R h_L)], \label{eqn2.13a} \\
&j_{th}(R,T) \nonumber\\
&= N_0 D \int dE \;  E[M_{00} (\partial_R h_L) + Q h_T +M _{30} (\partial_R h_T)].  \label{eqn2.13b} 
\end{align}
\end{subequations}
Here $D$ is the diffusion coefficient, $N_0$ is the electronic density of states, and $h_L$ and $h_T$ are the longitudinal and transverse quasiparticle distribution functions.  The terms in the square brackets in the integrands in Eqs. (\ref{eqn2.13a}) and (\ref{eqn2.13b}) are the spectral charge and thermal currents respectively.   

\begin{figure*}
\includegraphics[width=13cm]{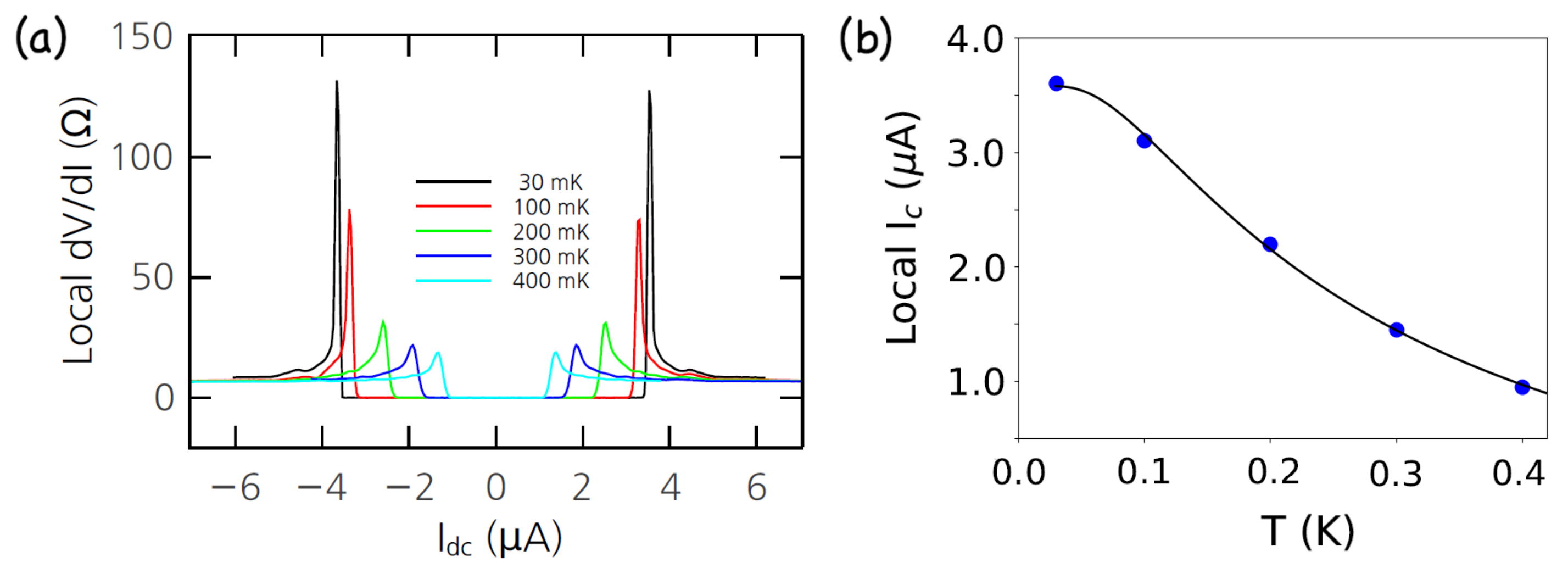}
\caption{(a) Local differential resistance of the device as a function of the dc bias current at 30, 100, 200, 300, and 400 mK. The critical current $I_c$ at the base temperature is $I_{c0}\sim$ 3.6 $\mu$A, which results in the factor $I_{c0} e R_N/E_c \sim$ 0.62. (b) Local critical current $I_c$ as a function of temperature. Solid line is a  fit to the functional form expected for a SNS junction in the long junction limit.}
\label{fig:Thouless}
\end{figure*}

\subsection{3. Boundary conditions}

The experimental geometries that we simulate consist of a network quasi-1D wires connected to each other, and to superconducting and normal metal contacts, which we model as `	reservoirs'' with well-defined values of the Green's function (and hence the Riccati parameters) at each reservoir.
On a normal reservoir, the Riccati parameters $\gamma$ and $\tilde{\gamma}$ are both zero.  On  a superconducting reservoir, they have the following values
\begin{subequations} \label{eqn3.2}
\begin{align}
\gamma_0(E) &= \frac{\Delta_0}{E + \sqrt{E^2 - |\Delta_0|^2}} \label{eqn3.2a} \\ 
\tilde{\gamma}_0(E) &= - \frac{\Delta_0^*}{E + \sqrt{E^2 - |\Delta_0|^2}}.  \label{eqn3.2b}
\end{align}
\end{subequations}
where $\Delta_0$ is the complex gap in the superconductor.  A magnetic flux is introduced by specifying the gauge-invariant phase $\phi$ of this parameter, i.e., $\Delta_0 = |\Delta_0| e^{i \phi}$.   
The distribution functions have the following equilibrium form in a superconducting or normal reservoir at a potential $V$:
\begin{equation}
h_{L,T}=\frac{1}{2}\left[\tanh\left(\frac{E+eV}{2 k_B T}\right) \pm \tanh\left(\frac{E-eV}{2 k_B T}\right)\right].
\label{eqn3.4}
\end{equation}

Typically, the interface between a normal wire and a superconducting reservoir will not be perfectly transparent.
The boundary conditions of Kupriyanov and Lukichev are often used \cite{kupriyanov}, but these are only valid in the tunneling limit, i.e., for small barrier transparency, although they also work for perfectly transparent barriers.  For arbitrary barrier transparencies, Nazarov \cite{nazarov2} has given a more general formula in terms of an interface with $N$ conducting channels, each with an arbitrary transmission coefficient $T_n$   
\begin{equation}
\hat{g}_{s1} \partial_x \hat{g}_{s1}=\alpha \frac{e^2}{\pi} \underset{n}{\sum}  \;2 T_n \frac{[\hat{g}_{s1},\hat{g}_{s2}]}{4 + T_n(\hat{g}_{s1}\hat{g}_{s2} + \hat{g}_{s2}\hat{g}_{s1}- 2)}.
\label{eqn3.5}
\end{equation}
Here, $\hat{g}_{s1,2}$ are the Green's functions on either side of the barrier, and $\alpha$ is a constant factor.  Obviously, since we do not know the individual transmission coefficients, this equation is difficult to use in its current form.  The open-source code that we have used assumes the simplest case of perfectly transparent interfaces between the 1D wires and the superconducting and normal reservoirs. 

\subsection{4. Solution procedure}

To obtain a solution, the differential equations for the Riccati parameters [Eq. \ref{eqn2.10}] are first solved with the boundary conditions at the normal metal and superconducting reservoirs specified as a function of energy $E$.  The gauge invariant phase $\phi_1$ across one SNS junction is used as a fitting parameter, with the phase $\phi_2$ across the second junction being obtained from the usual SQUID relation $\phi_2 = \phi_1 + 2 \pi \Phi/\Phi_0$. At nodes joining multiple 1D wires, the boundary conditions are continuity of the Riccati parameters, and a Kirchoff law for their derivatives, e,g., $\sum \partial \gamma = 0$, where the sum is over all the 1D wires joined at a node.  Once the Riccati parameters are obtained, the spectral currents can be calculated from Eqs. (\ref{eqn2.11}) and (\ref{eqn2.12}).  With this information, the kinetic equations can be solved using the boundary conditions (\ref{eqn3.4}) for the distribution function, and conservation of spectral currents at each node.

To determine the nonlocal voltage at a specific temperature and flux, the entire solution procedure described above is integrated into a numerical solver using the nonlocal voltage $V_{nl}$ and the phase $\phi_1$ as fitting parameters.  The conditions for the solver are that the net current into the voltage probe $V^+$ in Fig. 1(b) vanishes, and that the net current into the second superconductor [the one with the $V^-$ contact in Fig 1(b)] from both normal metals attached to it also vanishes.  Once the solver converges, we can then calculate the current $I_b$ through the current injection contact, the nonlocal differential resistance $dV_{nl}/dI_b$, as well as the supercurrents and quasiparticle currents through any 1D wire.   

$E_c$ is nominally determined by the length $L$ of the normal part of the SNS junction, $E_c = \hbar D/k_BT$.  However, this is for a wire with no additional normal metal leads.  Experimentally, by measuring the saturation value of $I_c$ at low temperatures, we have found that $E_c$ is reduced by a factor of about 20 from its expected value based on $L$ \cite{noh,noh2}.  This can be thought of as an increase in the effective value of $L$ due to the increased probability of quasiparticle diffusion in the leads.  Consequently, while $E_c \sim 55$ $\mu$eV for a length $L=$ 450 nm and $D=170$ cm$^2$/s, we have used a value of 2.7 $\mu$eV, adjusting the values of $\Delta$ and $T$ which are specified in units of $E_c$ accordingly.

\section*{Appendix B: Characterization of device}

\renewcommand{\thesection}{B} 
\setcounter{equation}{0}

Figure \ref{fig:Thouless}(a) shows the local differential resistances $R_{51,69}$ of the device presented in the main text at various temperatures. The critical current $I_c$ is plotted in (b) as a function of temperature, which shows a good agreement with the functional form expected for a SNS junction in the long junction limit \cite{dubos}. However, while the local critical current $I_c$ of a simple diffusive SNS junction consisting of 1D normal metal wire between two superconducting reservoirs in the long junction at the lowest temperature is given by 
\begin{equation}
I_c = 10.82 \left( \frac{E_c}{eR_N}   \right),
\end{equation}
this relation is no longer correct in the multiterminal case, where the multiple normal leads connected to normal contacts increase the effective length of the device, and consequently reduce the effective $E_c$.  This was already mentioned in Ref.  [\citenum{noh}] for linear devices.  For the device studied for this paper,  the maximum value of the critical current $I_{c0}$ of 3.6 $\mu$A corresponds to $I_{c0} e R_N/E_c \sim 0.62$ if we used the length $L$ of the normal metal between the two superconductors to calculate $E_c$. Detailed numerical calculations showing that the additional leads connected to the normal metal wire can result in a reduction of $I_{c0}$ can be found in Ref. \cite{noh2}.

\section*{Appendix C: Alternate device geometries}

\begin{figure*}
\includegraphics[width=13cm]{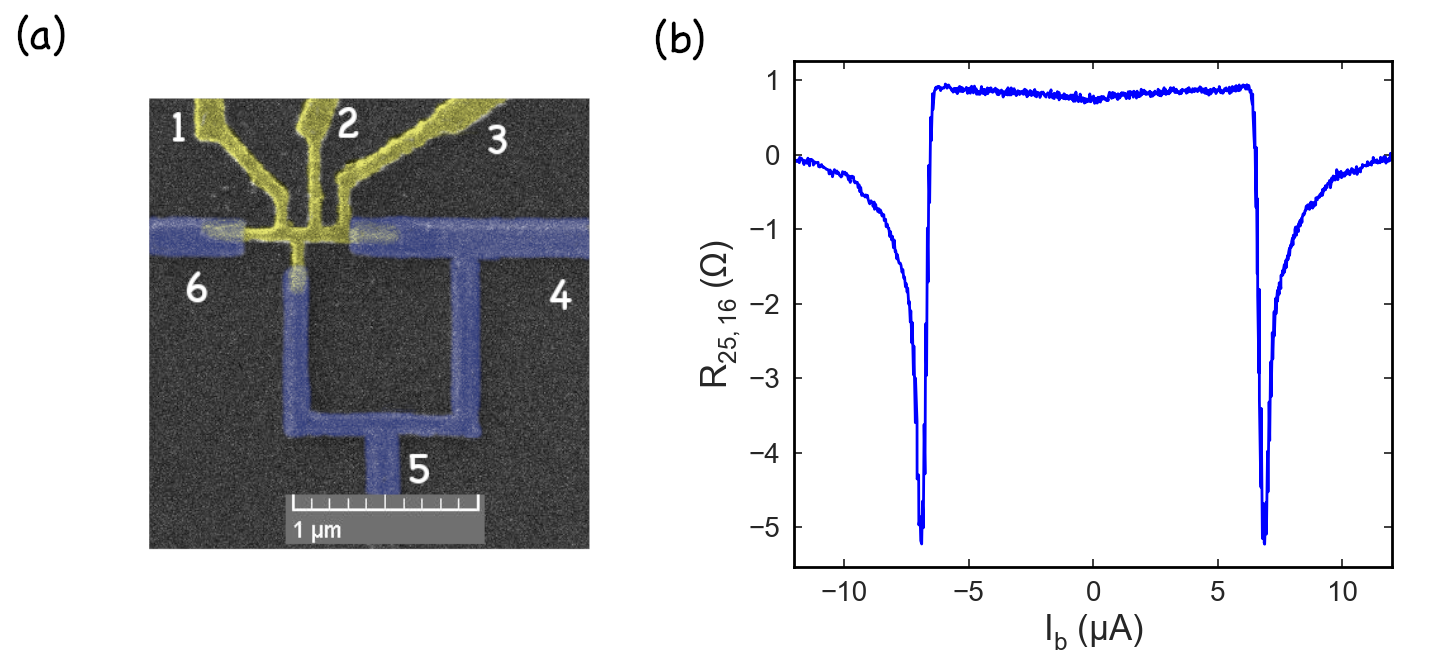}
\caption{(a)  False color SEM image of another device that we fabricated and measured.  Yellow represents the normal metal (Au) and blue the superconductor (Al).  (b)  Nonlocal differential resistance of the device in (a) taken at 26 mK as a function of the dc bias current $I_b$ sourced through contact 2 and drained through contact 5.  Numbers refer to contacts in (a).  For notation, see main text. }
\label{figs1}
\end{figure*}

Figure \ref{figs1}(a) shows a false color SEM image of one of the other device geometries that we fabricated and measured.  This device has three NS interfaces in contrast to the four NS interfaces in the device discussed in the main text.  Nevertheless, it shows similar nonlocal behavior.    Figure \ref{figs1}(b) shows the nonlocal resistance  $R_{25,16}= dV_{16}/dI_{25}$ as a function of the dc bias current $I_b$ sourced in contact 2 and drained from contact 5.  While the critical current is larger than the device in the main text, the response is nearly identical.


\end{document}